\documentclass[twocolumn,preprintnumbers,amsmath,amssymb,superscriptaddress]{revtex4}

\usepackage{graphicx}% Include figure files
\usepackage{dcolumn}% Align table columns on decimal point
\usepackage{bm}% bold math

\begin{document}

%\preprint{}

\title{A model for hand-over-hand motion of molecular motors}

\author{J. Mun\'{a}rriz}

\affiliation{Dpto. de F\'{\i}sica de la Materia Condensada,
Universidad de Zaragoza. 50009 Zaragoza, Spain}

\affiliation{Instituto de Biocomputaci\'on y F\'{\i}sica de Sistemas
Complejos, Universidad de Zaragoza. 50009 Zaragoza, Spain}

\author{J.~J. Mazo}

\affiliation{Dpto. de F\'{\i}sica de la Materia Condensada,
Universidad de Zaragoza. 50009 Zaragoza, Spain}

\affiliation{Instituto de Ciencia de Materiales de Arag\'on,
C.S.I.C.-Universidad de Zaragoza. 50009 Zaragoza, Spain.}

\author{F. Falo}

\affiliation{Dpto. de F\'{\i}sica de la Materia Condensada,
Universidad de Zaragoza. 50009 Zaragoza, Spain}

\affiliation{Instituto de Biocomputaci\'on y F\'{\i}sica de Sistemas
Complejos, Universidad de Zaragoza. 50009 Zaragoza, Spain}

\date{\today}

\begin{abstract}
A simple flashing ratchet model in two dimensions is proposed to
simulate the hand-over-hand motion of two head molecular motors like
kinesin. Extensive Langevin simulations of the model are
performed. Good qualitative agreement with the expected behavior is
observed. We discuss different regimes of motion and efficiency
depending of model parameters.
\end{abstract}

\maketitle

\section{Introduction}

The dynamics of molecular motors is an important topic in biophysics
and nanotechnology. In the living and in the artificial nanoscale
world fast non-diffusive directed transport or rotary motion
constitute key ingredients of any complex structure. Molecular motors
are the ``nanomachines'' which perform these tasks
\cite{Schliwa,Dekker, Vale,Howard}. This definition involves a
considerable amount of different molecules: Motor proteins, such as
myosin and kinesin, RNA polymerases, topoisomerases, ...

In this paper we focus on the problem of directed motion over a
substrate which is exemplified in the kinesin \cite{Carter,Asbury2}.
Active transport in eukariotic cells is driven by complex proteins
like kinesin which moves cargo inside cells away from the nucleus
along microtubules transforming chemical fuel (ATP molecule) into
mechanical work.  Kinesin is a two head protein linked by a domain
(neck) and a tail which attach a cargo or vesicle to be carried.  The
two heads perform a processive walk over the substrate (the
microtubule).  The way in which this process is performed attracts big
interest in the research in molecular biology as well as in biological
physics. In order to understand how kinesin works two properties that
arise from the structure \cite{julicher, Howard} of the microtubules
cannot be forgotten: they have a regular, periodic structure and
structural polarity -- they are asymmetric with respect to their two
ends, which determines the direction of kinesin motion

\begin{figure}
\includegraphics[width=8.5cm]{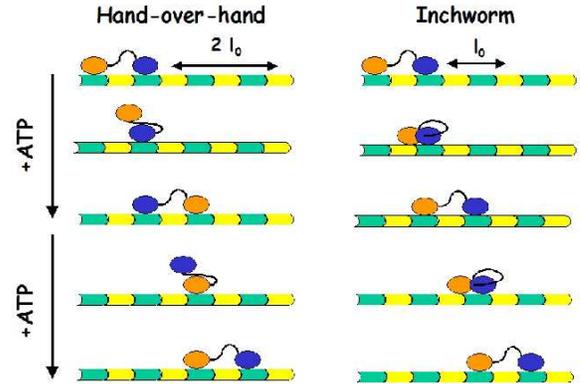}
\caption{\label{fig:hand_inch} (Color online) Schematic representation
of the possible mechanisms of motion for the motor of kinesin. In the
hand-over-hand case each head moves a distance equals to $2l_0$
whereas in the inchworm the period of motion is $l_0$.}
\end{figure}

In the last fifteen years, experimental molecular biology has provide
a lot of new results which allows to elucidate, at mesoscopic level,
the main mechanisms for directed transport. These experimental
evidences are mostly based in single molecules
experiments~\cite{Visscher, Ritort}. The interpretation of these
results is not always easy and many times are not conclusive on the
detailed way in which the motor walks.  Two basic mechanisms have been
proposed to explain the kinesin motion: ``inchworm'' and
``hand-over-hand'' motion (see figure \ref{fig:hand_inch}). In the
first case, one head does not overcome the other one. In this case the
period of the motion is one period of the microtubule structure ($l_0$
in the figure). In the hand-over-hand mechanism one head overcomes the
other. Now the period for each head is the double ($2l_0$). In both
cases the center of masses advances the same length. Although first
single molecule experiments were compatible with both mechanisms more
recent experiments have shown \cite{Yildiz,Asbury,Schief,Hua} in a
very clever way that hand-over-hand motion may be more plausible.

Two strategies can be devised in order to model the motion of
molecular motors~\cite{Fisher,Chow}.  On one hand continuous models
based on mirror symmetry breaking potentials (ratchet
potentials~\cite{Reimann}) or time symmetry broken driven
forces~\cite{Chacon}.  On the other hand, discrete kinetics models
which are based on the solution of master equations associated to
different states of the motor (see~\cite{Fisher} and references
therein).  Using either approximation both mechanisms have been
studied: inchworm~\cite{Cilla,Sancho2} or
hand-over-hand~\cite{SaoGao,Ping,Sancho1,Sasaki}.

In this work we study a minimalist mechanical continuous model for
hand-over-hand motion that, we believe, captures the main features of
biological motors. The model also has into account the properties of
the microtubule substrate. The article is organised as follows: first
we cast a two dimensional model which can mimic the motion of the
motor. Within of reasonable values of parameters we explore different
regimes of motion.  In the conclusions section we will discuss the
validity of the results to model a molecular motor.

\section{2-D MODEL}

In order to find a suitable model for the kinesin motor, its
properties shall be studied carefully. Ref. \cite{Carter} summarizes
all these features: kinesin is a two-head protein which moves along
the microtubule with $8.3\,nm$ steps, matching the repeat distance of
the microtubule lattice; each step needs 1 ATP which is hydrolyzed and
the movement stalls when a backward load of $7\,pN$ is
applied. Experiments reported in \cite{Yildiz} show, by marking one of
the heads, that the motion follows the hand-over-hand mechanism,
as a $16.6\,nm$ step is observed for each head, thus forbidding the
movement proposed in the inchworm mechanism.

The description of the movement is rather simple: the two heads of the
kinesin are attached to the microtubule \cite{Mori} in two neighbor
monomers until 1 ATP molecule is hydrolyzed by the head
backwards. This energy frees the head, which moves to a new binding
place ahead the other one. Two complementary mechanisms to understand
how the particle released is able to find the next binding site have
been proposed~\cite{Carter,Tomi}: (a) The \emph{neck linker} mechanism
assumes a conformational change in the neck between heads which moves
the free head from one place to the next forwards. (b) The
\emph{diffusional search} relies on the assumption that the noise
associated to the thermal bath that surrounds the particle makes the
free particle move, and this movement is preferably forwards and
forced by the particle ahead which is attached to the microtubule.

Thermal fluctuations play a central role in the whole process. In the
nanometer-length dimension and at room temperature, motion is governed
by randomness induced by the environment (in this case the cytosol,
made up mainly by water). At this scale damping and thermal noise are
dominant and the dynamics can be studied by an overdamped Langevin
equation:
\begin{equation}
\gamma \frac{d{\bf r}}{dt} = -{\bf \nabla} V({\bf r}) + {\bf F}
({\bf r},t)+{\boldsymbol \xi}(t).
\label{eq:overdampedlangevin}
\end{equation}
Here ${\bf F}$ stands for external forces and ${\boldsymbol \xi}$ for thermal
noise, being
\begin{equation}
< \xi_j(t)\xi_k(t') > = 2\,\gamma\,k_B\,T\, \delta(t-t') \delta_{jk}
\label{eq:autocorrelacion}
\end{equation}
($\xi_j$ and $\xi_k$ are Cartesian components of the vector
${\boldsymbol \xi}$).

\subsection{Energy potentials}

We will model the kinesin as two interacting particles moving in the
plane under the effect of flashing ratchet substrate potentials (two
particles moving in two dimensions).

The potential energy of the system is given by
\begin{equation}
V({\bf r_1},{\bf r_2})=V_1({\bf r_1},t)+V_2({\bf r_2},t)+V_{12}({\bf
r_1}-{\bf r_2})
\end{equation}

The two heads of the kinesin are linked through a modified version of
the Finite Extensible Non-linear Elastic (FENE) interaction
\cite{FENE}:
\begin{equation}
V_{12}(r)=-\,\frac{1}{2}K\,R_0^2\,\log \left( 1 - \frac{(r-l_0)^2}{R_0^2} \right),
\label{FENEeq}
\end{equation}
with $r=|{\bf r_1}-{\bf r_2}|$, $K$ is the stiffness of the neck,
$l_0$ the equilibrium distance between heads and $R_0$ determines a
maximum allowed separation, $l_0-R_0<r<l_0+R_0$.

With respect to the substrate potentials, in order to model the
characteristics observed, two periodic flashing ratchet potentials
lagged half a period in the $x$ direction will be used.
\begin{equation}
V_j({\bf r},t)=V_j({\bf r})f_j(t).
\end{equation}
$j=1,2$ and in the $x$ direction the potentials are periodic with
period $2l_0$ and $V_2$ is displaced $l_0$, the period of the
microtubule lattice~\cite{l0}, with respect to $V_1$:
\begin{equation}
V_1({\bf r}+2l_0 {\bf \hat{x}})=V_1({\bf r})=V_2({\bf r}+l_0 {\bf \hat{x}}).
\label{period}
\end{equation}

\begin{figure}[]
\includegraphics[width=8.5cm]{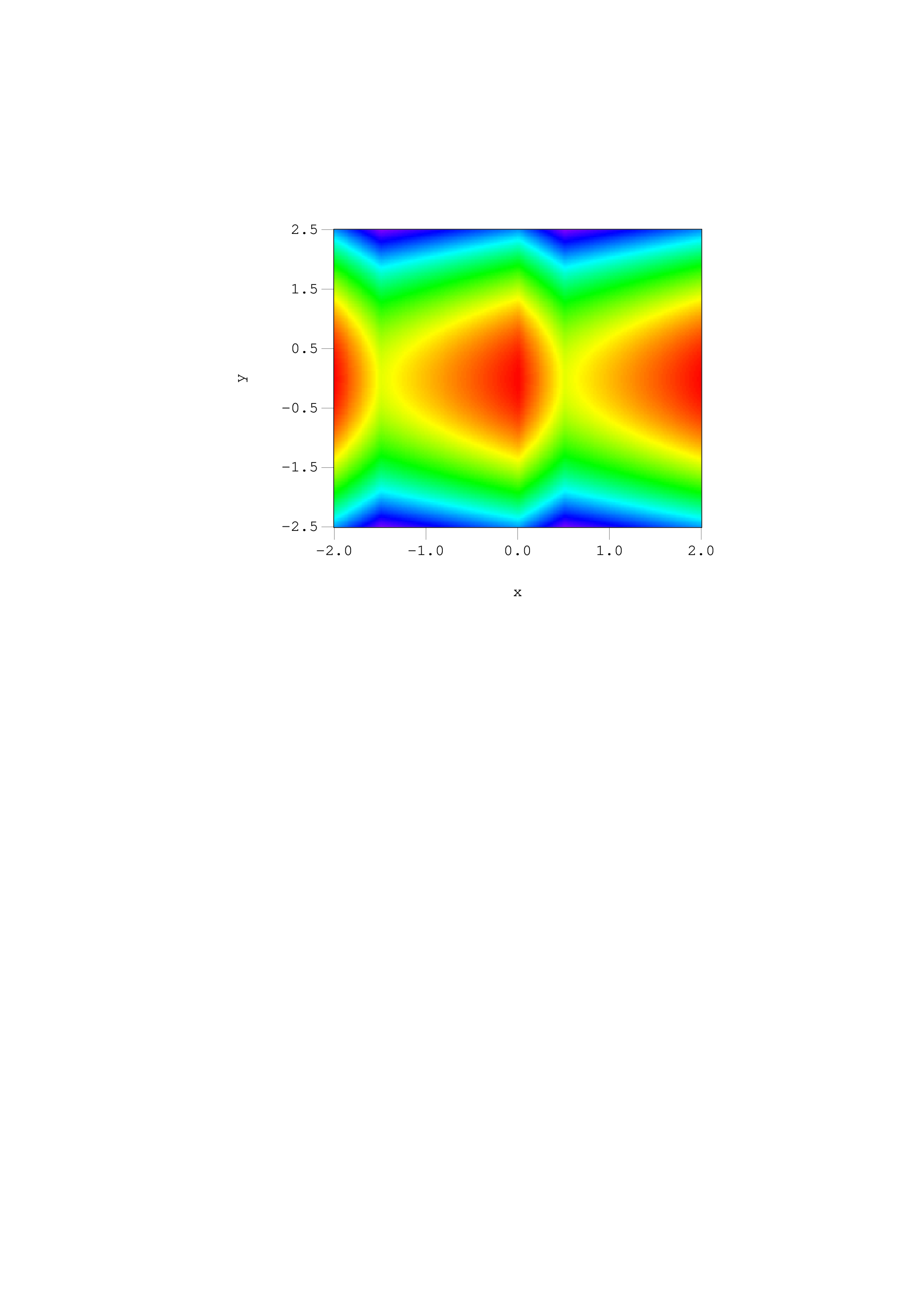}
\caption{(Color online) Surface plot of the 2d substrate potential
$V_1$. Minima correspond to $\widetilde{x}=2n$ ($n=0,\pm 1,...$) and
$\widetilde{y}=0$.}
\label{fig:potential} 
\end{figure}

The mathematical description of the 2d potential associated to
particle 1 [see Fig.~(\ref{fig:potential})] is the following
\begin{equation}
V_1(x,y) = V_{1x}(x) + V_{1y}(y)
\end{equation}
with
\begin{equation}
V_{1x}(x)=\left\{ \begin{array}{clc} \frac{x}{x_M}V_0 & {\rm if} & 0
\leq x \leq x_M \\ 
\\ \frac{2l_0-x}{2l_0-x_M}V_0 & {\rm if} & x_M \leq x
\leq 2l_0. \\
\end{array}
\right.
\end{equation}
$x_M$ controls the asymmetry of the potential and if $x_M=l_0$ the
potential is symmetric.

In order to confine the particles in the microtubule channel, with
respect to the $y$ direction we choose a simple parabolic dependence.
\begin{equation}
V_{1y}(y)=\frac{1}{2}k_y\cdot y^2
\end{equation}

\begin{figure}[t]
\includegraphics[width=8cm]{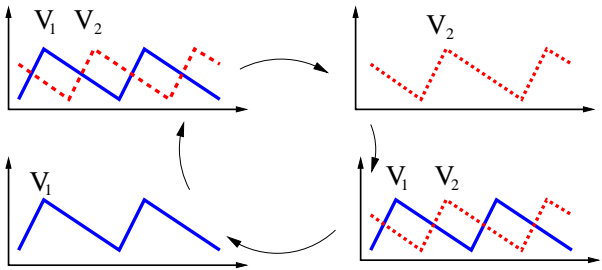}
\caption{(Color online) Time sequence for the flashing ratchet
  substrate potentials. Each potential acts on a different
  particle. Note that this potential follows the sequence of
  attach-deattach shown in the hand-over-hand motion,
  Fig.~\ref{fig:hand_inch}.}
\label{fig:time} 
\end{figure}

We still have to define $f_j(t)$. The idea is to reproduce a cyclic
motion. Such a cycle has 4 steps, see Fig.~(\ref{fig:time}. First
($t=0$) both particles are confined close to the minima of their
respective potentials and thus separated an averaged distance $l_0$
(the natural length of the neck). After a given time $t_{\rm on}$ some
energy arrives at particle 1 for instance which does not see its
substrate potential for a time $t_ {\rm off}$. During this time this
particle suffer a thermal diffusion only subjected to the interaction
with the other particle. When $V_1$ is switched on again at $t=t_{\rm
on}+t_{\rm off}$, the particle slides down towards some minimum energy
position. This step lasts another $t_{\rm on}$ time and then at
$t=2t_{\rm on}+t_{\rm off}$, $V_2$ is switched off for a $t_{\rm off}$
time closing the cycle.  The total period of this cycle is $T=2t_{\rm
on}+2t_{\rm off}$. As we will see, thanks to the asymmetric character
of the potential a directed motion is obtained.

In order to compute the efficiency of the motion we define an efficiency
parameter given by:
\begin{equation}
\varepsilon = \frac{\langle \Delta x_1 \rangle}{2l_0} \times 100,
\end{equation}
where $\langle \Delta x_1 \rangle$ is the average advance of particle
1 (for instance) per cycle of the potential. Note that our definition
of efficiency is basically the velocity of the motion, in fact the
mean velocity can be computed as $v_{\rm mean}=\frac{\epsilon}{100}
\times 2l_0/(t_{on}+t_{off})$ and is not related to the input of
energy and the output of work.

The important parameter here is $t_{\rm off}$, the time a particle has
for the diffusive motion. $t_{\rm on}$ only requires to be long enough
to allow for relaxation towards a minimum, which in overdamped
dynamics happens very fast. Thus, in our simulations we have played
with different values of $t_{\rm off}$ and set $t_{\rm on}=t_{\rm
off}$. This value corresponds to a duty ratio $r=t_{\rm on}/(t_{\rm
on}+t_{\rm off})=0.5$ which guarantees the processivity of the
motion~\cite{Howard,Chow}.

\subsection{\label{ss:norm}Normalization}

We will measure distance in units of $l_0=8.3$\,nm, the distance
between monomers in the microtubule, see also~\cite{l0}. Energy is measured
in units of $V_0$, the maximum value of the substrate potential. We
choose $V_0 \simeq E_{\rm ATP} \simeq 20$ $k_BT$ (at 300\,K)
\cite{com1}. The natural unit of time will be $\tau=l_0^2 \gamma / V_0
\simeq 40$\,ns. Here, $\gamma$ is the damping coefficient used in the
Langevin equation ($\gamma=6\pi\eta r= 4.7 \cdot 10^{-11}
\frac{kg}{s}$, with $\eta=10^{-3}$\,Pa the viscosity of the water and
$r=25\,\AA$ the size of the head).

%If we write the normalised dynamical equations for one of the particles in
%the $x$ direction we get:
%\begin{equation}
%\frac{d\widetilde{x}}{d\widetilde{t}} = - \frac{\partial
%\widetilde{V}}{\partial \widetilde{x}} + \widetilde{Q} +
%\widetilde{\xi}(\widetilde{t})
%\label{eq:dim_less}
%\end{equation}
%with
%\begin{equation}
%\langle \widetilde{\xi}(\widetilde{t}) \widetilde{\xi}(\widetilde{t'}) \rangle =
%2\,\widetilde{T} \delta(\widetilde{t}-\widetilde{t'}).
%\label{eq:autoco}
%\end{equation}

We will use now \;$\widetilde{}$\; signs for normalized variables as
\begin{eqnarray}
\widetilde{x}=\frac{x}{l_0} \qquad ; \qquad \widetilde{t}=\frac{t}{\tau} \quad
; \qquad \widetilde{V}=\frac{V}{V_0} \nonumber \\
\widetilde{T}=\frac{k_B\,T}{V_0} \qquad {\rm and} \qquad \widetilde{Q}=\frac{l_0\, Q}{V_0}
\label{eq:normalizaciones}
\end{eqnarray}

\begin{figure}
\begin{tabular}{c}
\includegraphics[width=8.5cm]{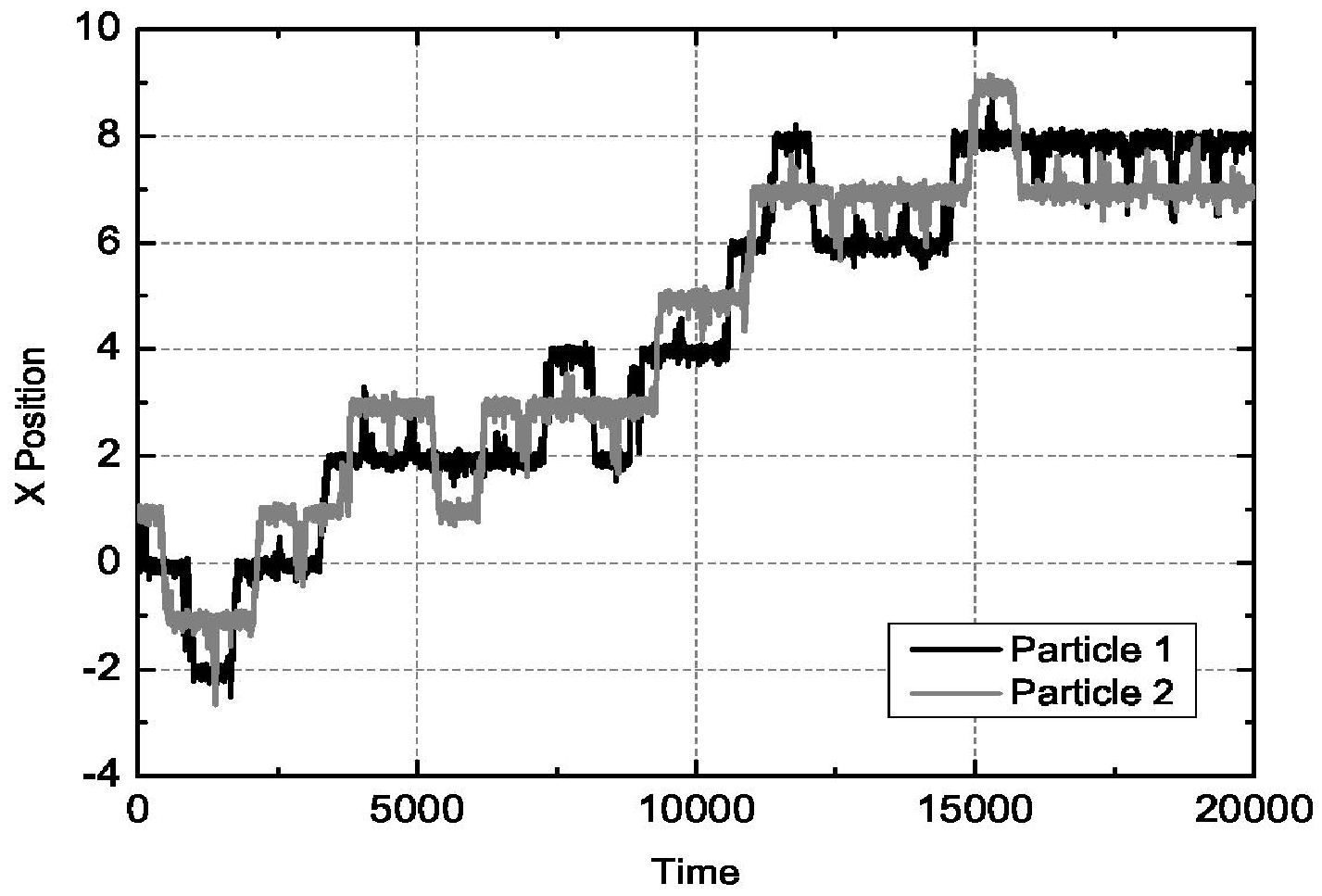}\\\includegraphics[width=8.5cm]{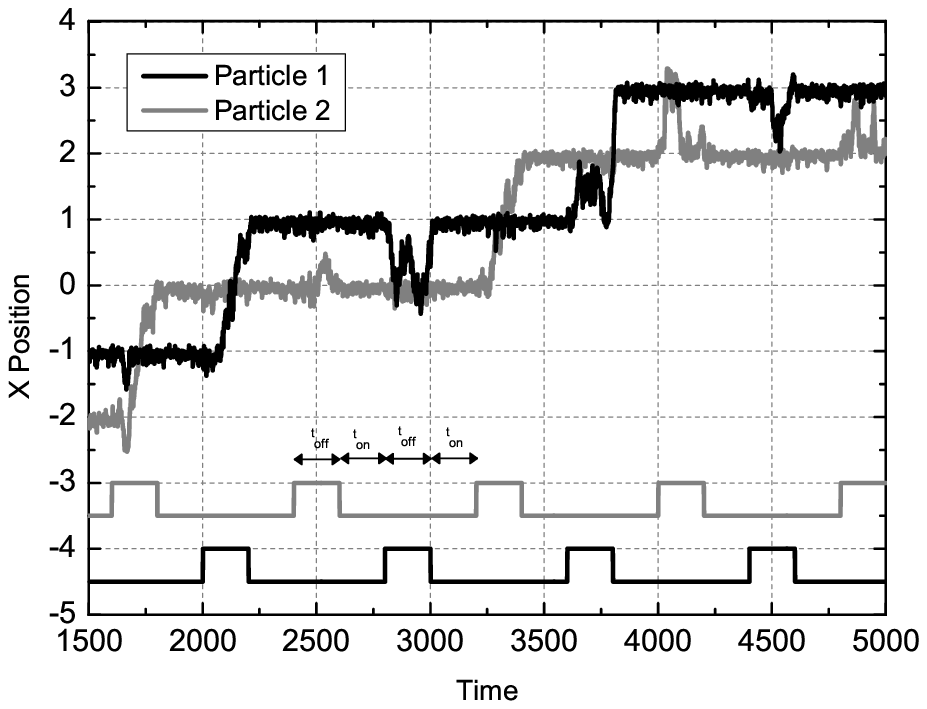}\\\includegraphics[width=8.5cm]{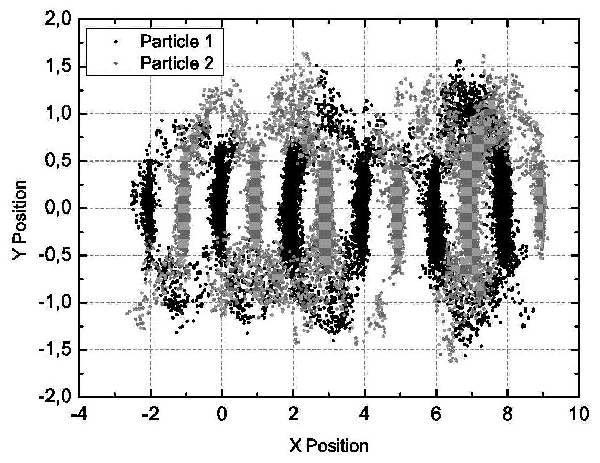}\\
\end{tabular}
\caption{\label{fig:dynamics} $\widetilde{x}$($\widetilde{t}$) for the two
particles (top and middle) and trajectory in the $\widetilde{x}$--$\widetilde{y}$
plane (bottom). The middle figure also shows the flashing dynamics of the
substrate potentials (the base lines correspond to the {\em on} periods).}
\end{figure}

\section{\label{sec:results} RESULTS}

We are going to present our results based in the numerical integration
of the normalised system of equations for the two particles. The
integration algorithm we use is a version of the Runge-Kutta algorithm
for integration of stochastic differential equations ($3_O\,4_S\,2_G$)
\cite{sde1,sde2}. With respect to the different constants and
parameters, unless extra information is given, the default normalized
parameters will be we $\widetilde{T}=0.05$ (300 K), $\widetilde{t}_
{\rm off}=\widetilde{t}_ {\rm on}=20$, $\widetilde{K}=10$,
$\widetilde{k}_y=1$, $\widetilde{R}_0=0.4$ and $\widetilde{x}_M=0.5$.

\subsection{\label{sub:dynamics} Dynamics of the system}

Fig.~\ref{fig:dynamics} shows a typical example of the dynamics of the
system at the parameter values listed above. There we can see that
simulations reproduce the expected mechanism, a hand-over-hand net
advance of the molecule. Middle figure shows a detail of the top one.

If both potentials are on, the particles do random motions around the
minimum potential energy position. However, as one of the potentials
is turned off, its linked particle starts to diffuse in 2d. The
importance of the asymmetric mechanism is fully understood here. After
$t_ {\rm off}$, when the potential is turned on again, most of the times
the particle is sited to the right of the maximum of the asymmetric
potential and then typically moves down to the nearest minimum
position. As we have said, due to the asymmetry of the potential, this
minimum more frequently corresponds to the one to the right of the
original one. Clearly, the more asymmetric the potential is, the more
likely the system moves forward.

In the third graph of Fig.~\ref{fig:dynamics} we show the trajectories
of the particles in phase space. The distance between heads moves
around the rest distance $l_0$. Motion in the $x$ direction happens
usually when one of the substrate potentials is off. Otherwise
particles stay most of the time close to minimum energy position.

\subsection{Efficiency as a function of $t_{\rm off}$ and $T$}

\begin{figure}
\includegraphics[width=8.5cm]{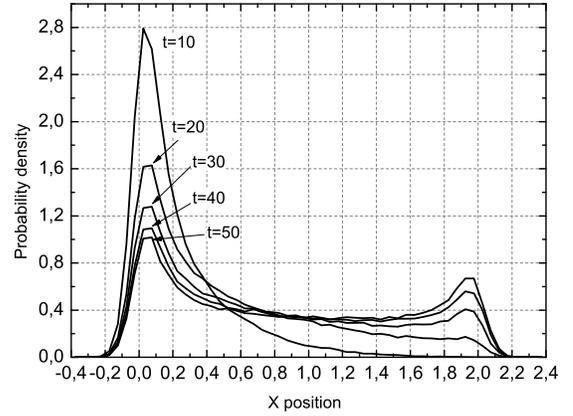}
\caption{\label{fig:histo_x} x-axis projection, at different
normalized times $\widetilde{t}$, of the diffusion of a particle attached
to the other, fixed at $(1,0)$, when no substrate potential is being
applied. Data obtained at $\widetilde{T}=0.05$}
\end{figure}

A first estimation of the time needed for a particle to reach the next 
minimum can be easily worked out by using the 2D diffusion equation for the 
particle probability distribution $p(\varphi,t)$,
\begin{equation}
\label{eq:diffusion}
\frac{\partial p}{\partial t} = D \, \nabla^2 p.
\end{equation}

Writing [\ref {eq:diffusion}] in polar coordinates and assuming that
the distance between heads $r$ is constant the equation reads
\begin{equation}
\label{eq:polar}
\frac{\partial p}{\partial t} = \frac{D}{r^2}\,\left[ \frac{\partial^2
p}{\partial r^2} + \frac{1}{r}\,\frac{\partial p}{\partial r} +
\frac{\partial^2 p}{\partial \varphi^2} \right]_{r=const.} =
\frac{D}{r^2}\frac{\partial^2 p}{\partial \varphi^2}
\end{equation}
This equation can be solved (making Fourier transformation for
instance) with appropriated initial conditions
\begin{equation}
\label{eq:ini_cond}
\left. p(\varphi,t) \right|_{t=0} = \delta (\varphi)
\end{equation}
to give the normalized $p(\varphi,t)$
\begin{equation}
\label{eq:sol_dif}
p(\varphi,t)=\frac{r}{2\sqrt{\pi\,D\,t}} \exp\left( \frac{-r^2
\varphi^2}{4\,D\,t} \right)
\end{equation}
from this results the mean angle reached at time $t$ is given by 
\begin{equation}
\langle \varphi^2 \rangle = \frac{2 D t }{r^2} = \frac{2 k_B\,T t }{\gamma r^2}
\end{equation}
where we have used the Stokes-Einstein relation:
\begin{equation}
D=\frac{k_B\,T}{\gamma}.
\end{equation}

Let be $\varphi_M$ the angle where the maximum of the potential is
placed which is determined by the $x$ position of that maximum,
$x_M$. If the particle is at a $x$ position less than $x_M$ when the
potential turns on, it will return to its original position. However,
if $x>x_M$ the particle will move forward. Assuming that the position
of the maximum is placed at $\widetilde{x}_M=0.5$, and
$\widetilde{r}=1$ we obtain $\varphi_M=\pi/3$.  Figure
\ref{fig:histo_x} shows the time evolution of the probability
distribution (projected on the x-axis). It is clearly observed that as
time goes the probability of finding that a particle crosses the maximum
$x_M$ increases. The time in which the probability for crossing is
$1/2$ is simply given by
\begin{equation}
\label{eq:t_half}
t = 0.674 \frac{r^2 \varphi_M^2 \gamma}{2 k_B\,T}.
\end{equation}

With the values given above the adimensional time (for temperature
$\widetilde{T}=0.05$) is $\widetilde{t}\sim 16$. For this $t_{\rm off}$ time
the efficiency of the motor is half of the maximum one, which is fixed
by $x_M$ (see below).

\begin{figure}[t]
\includegraphics[width=8.5cm]{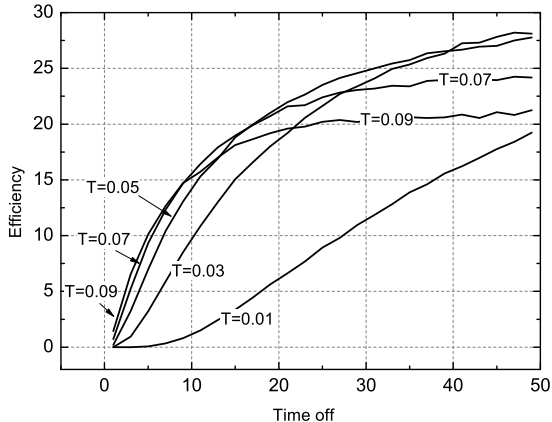}
\caption{\label{fig:T_t_int} Efficiency as a function of
$\widetilde{t}_{\rm off}$ for different normalized temperatures
$\widetilde{T}$.}
\end{figure}

Finally we analyze the behavior of the efficiency with temperature,
Fig.~\ref{fig:T_t_int}. For low temperatures we need long $t_{\rm
off}$ times to reach a reasonable efficiency as expected from equation
[\ref{eq:t_half}] and we not get an asymptotic limit. For intermediate
temperatures $\widetilde{T}=0.03-0.05$ the highest efficiency is
achieved. Moreover, in the limit of high temperatures compared to the
2d potential and long $t_{\rm off}$, the efficiency starts to fall, as
backwards movement is more likely to occur (the free particle can drag
the confined one).

\begin{figure}
\includegraphics[width=8.5cm]{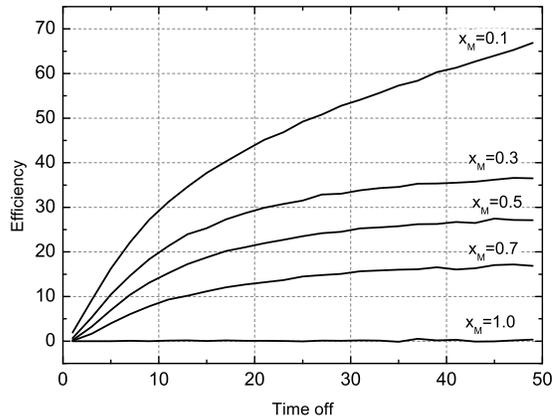}
\caption{
\label{fig:x_m_t_int} Efficiency as a function of $\widetilde{t}_{\rm off}$ at
different positions of the maximum $\widetilde{x}_M$. When
$\widetilde{x}_M=1.0$, the potential becomes symmetric and no rectified
movement is observed.}
\end{figure}

\subsection{Efficiency at different asymmetries}

Here we present results on the behavior of the system as the asymmetry
of the potential changes, being $x_M$ the parameter that controls it
($x_M$ fixes the position of the maximum in the period $2l_0$ periodic
potential, so $\widetilde{x}_M=1$ corresponds to the symmetric
case). At a given $t_ {\rm off}$ time, efficiency depends importantly
on this parameter. The mechanism will be inefficient for a symmetric
potential and the largest efficiency will be obtain for the more
asymmetric one.

Fig. \ref{fig:x_m_t_int} shows the numerical simulation of the
efficiency as a function of $t_ {\rm off}$ for different values of
$x_M$.  As we reduce the asymmetry the efficiency tends to zero, as
shown in the $\widetilde{x}_M=1.0$ line, which corresponds to a symmetric
potential. On the other hand the efficiency of the mechanism increases
as we make the potential more asymmetric. In all the cases, when we
increase $t_{\rm off}$ the efficiency grows from zero and saturates at
its maximum value for long enough values of this parameter.

\begin{figure}
\includegraphics[width=8.5cm]{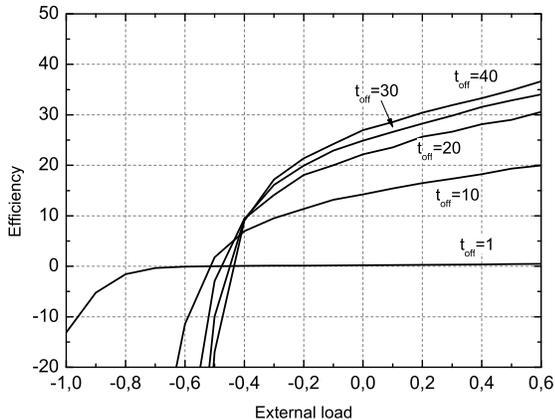}
\caption{\label{fig:carga_t_int} Efficiency as a function of the
external load applied, $\widetilde{Q}$. Each line refers to a
different $\widetilde{t}_{\rm off}$.}
\end{figure}

\begin{figure}
\includegraphics[width=8.5cm]{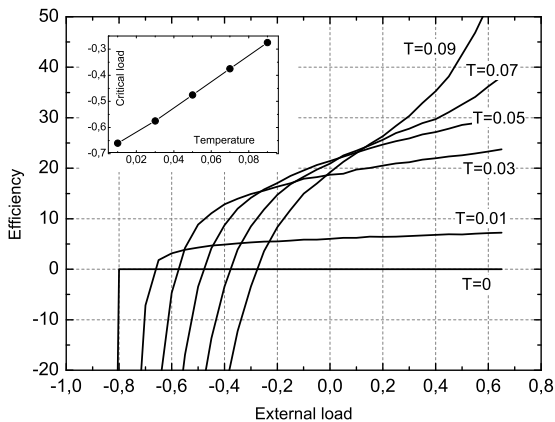}
\caption{\label{fig:carga_T} Efficiency as a function of the external
load applied $\widetilde{Q}$, for different normalized temperatures
$\widetilde{T}$.}
\end{figure}

\subsection{Dynamics under external loads}

In this section we want to explore the experimental results reported
in Ref.~\cite{Carter}, where backward stepping was observed when using
high backward loads. Then, it is worth studying how the system behaves
under the effect of an external force.

To model the effect of such a load is not trivial. We have to decide
how the total load $Q$ is divided into the two heads of the
protein. It seems obvious that a head can make an opposite force to
the applied only in case it is fixed to the
microtubule. Therefore, the following mechanism is proposed: if there
is only one head with its potential switched on, it will bear the
whole opposite load. On the other hand, if both heads have their
potential on, everyone will bear a force $Q/2$.

The expected behavior of the system is the following: as one potential
turns off, its associated head starts diffusing. The other particle
feels a force $Q$ which doubles the previous $Q/2$. Therefore, if that
force is strong enough. the particle starts climbing the potential
slope. The asymmetric potential plays again an important role: if the
external force is positive, the particle faces the sharpest slope of
the potential, so a bigger force than in the negative case is needed.

Fig.~\ref{fig:carga_t_int} shows the relationship between external
load and $t_{\rm off}$. The most important characteristic is the value
of the load for which the system does not move, 0 efficiency. For
negative loads, as $t_{\rm off}$ shortens, the particle needs greater
forces to start to move backwards. On the other hand, when long times
are employed, the mechanism seems to reach a limit around
$\widetilde{Q}=-0.5$.

We have also studied the effect of the temperature in the
mechanism. Results are shown in Fig.~\ref{fig:carga_T} where
efficiency versus external load for a given value of
$\widetilde{t}_{\rm off}=20$ is plotted at different temperatures. An
almost linear relation between critical load and temperature is
obtained in this range.

\begin{figure}
\includegraphics[width=8.5cm]{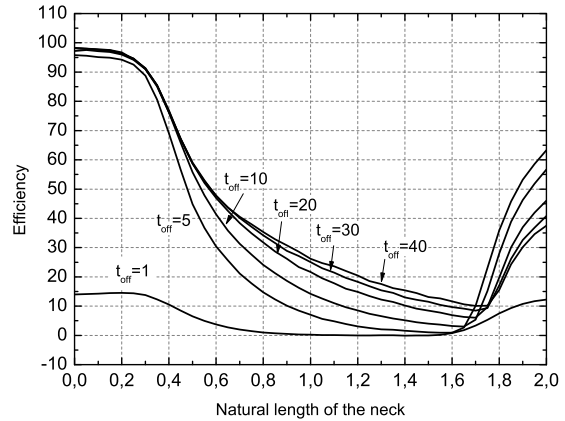}
\caption{\label{fig:l0_t_int} Efficiency as a function of the natural
length of the neck, $\widetilde{l}_0'$, for different values of $\widetilde{t}_{\rm
off}$.}
\end{figure}

\subsection{Varying the natural length of the neck}

Up to now we have studied the case where the two space lengths of the
system, the distance between monomers in the microtubule and the
natural length of the neck, are equal (both are $l_0$). In this
section we have extended our work to the study of the case when the
natural distance between the heads is different from the spatial unit,
fixed by the distance between monomers in the microtubule. Then, in
our model, $l_0$ need to be replaced by $l_0'$ in Eq.~(\ref{FENEeq}).

\begin{figure}
\includegraphics[width=5.5cm]{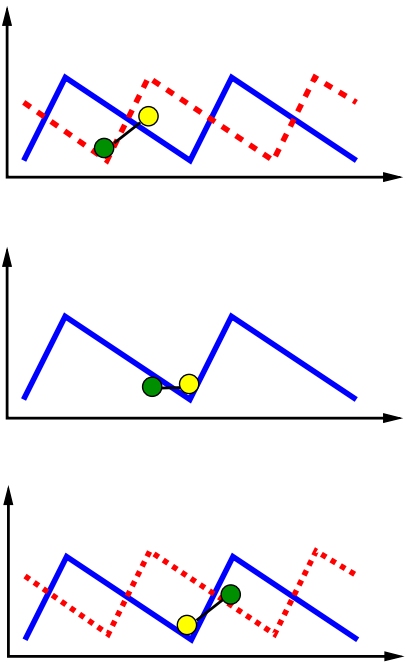}
\caption{\label{fig:l0_moves} (Color online) Schematic explanation of
the almost deterministic motion observed when $l_0'$ close to zero.}
\end{figure}

Fig.~\ref{fig:l0_t_int} is clear enough to provide strong evidence
about the striking behavior observed as the natural length of the neck
tends to 0: $100\%$ efficiency is achieved. This almost deterministic
mechanism can be understood with the help of Fig. \ref{fig:l0_moves}
and presents three steps: (a) We start with one particle sited in a
minimum of the potential and the other one ahead (it feels a small
force since the potential slope there is also small). (b) As the first
potential disappears, the second particle moves to its minimum,
dragging the other one. (c) Now the potential turns on, thus making
the first particle to move ahead and we recover a situation equivalent
to step (a).

\begin{figure}
\includegraphics[width=8.5cm]{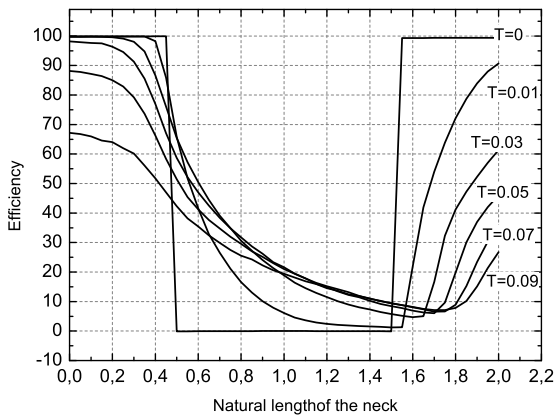}
\caption{\label{fig:l0_T} Efficiency versus $\widetilde{l}_0'$ for
different temperatures, $\widetilde{T}$.}
\end{figure}

\begin{figure}[t]
\includegraphics[width=8.5cm]{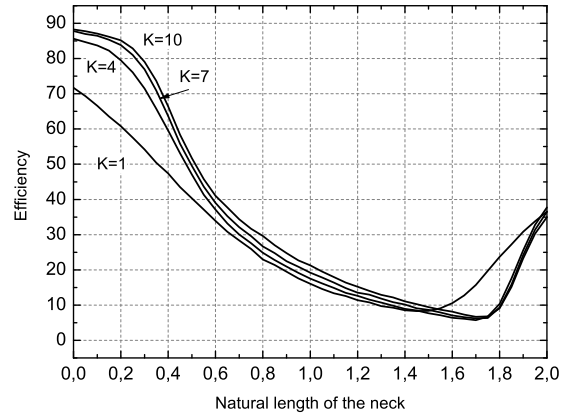}
\caption{\label{fig:l0_K} Efficiency versus $\widetilde{l}_0'$ for
different values of $\widetilde{K}$.}
\end{figure}

Fig. \ref{fig:l0_T} shows results at different temperatures. First of
all, the deterministic $T=0$ limit must be carefully explained. In
this limit, there are only two possible values for the efficiency:
$0\%$, associated to the range $\widetilde{l}_0'\in \left( 0.5,1.5
\right)$, and $100\%$ for $\widetilde{l}_0'\in \left[ 0.0,0.5 \right)\cup
\left( 1.5,2.0 \right]$. These two regions can be fully explained
using the mechanism described above. In the $0\%$ case, switching off
one of the potentials will make the other particle move to a minimum,
but it will not be the minimum ahead which would not produce a net
movement forwards.

There is just one parameter left to be discussed, which is the
stiffness of the linker between heads of the motor. The study of the
efficiency as a function of $l_0'$ at different values of $K$ is shown
in Fig. \ref{fig:l0_K}. For $\widetilde{l}_0' < 0.5$, the stiffness of the
neck determines whether the particles prefer to be in their minimums
no matter how far they are, or in an intermediate position, as plotted
in Fig. \ref{fig:l0_moves}~a).

\section{CONCLUDING REMARKS}

We have studied a simple mechanical model for hand-over-hand motion in
two dimensions. This model has into account some important
characteristics of two heads biological motor as kinesin. These
characteristics are incorporated in the model in a simple but
realistic way. The hand-over-hand mechanism requires a two-dimensional
space. Unidirectionality is given by the ratchet potential in the
advance direction. The balance between on and off times controls the
efficiency and processivity of the motion. With all these ingredients
we have been able to simulate the most remarkable features of kinesin
motion within reasonable values of he parameters.  Specifically, we
have clearly observed a stochastic directed motion in which particles
alternates each other (hand-over-hand). Moreover, a strong dependence
of the stall force with off time and temperature has been
found. Temperature makes a decrease of stall force with respect to one
expected from energetic calculations. This decrease in the motor
efficiency agrees with experimental observations \cite{Carter,
Fisher}.

Several improvements to the model can be considered in future work. A
link between $t_{\rm off}$ and ATP concentration could be
established. This would imply a random flashing force instead of the
periodic one used here. Another interesting extension of the model
could allow the motor to change the lane in $y$ axis. This could be
easily implemented by using a periodic potential in the transverse
direction.

Finally, we have to stress that the characterization of the behavior
and properties of those motors and the mechanisms behind them is an
initial step toward the construction of synthetic nanoscale motors.
This is a very active field in the nanoscience world. There has been
some successful achievements in this field that include triptycene
motors \cite{motor1}, helicene motors \cite{motor2} and a nanotube
nanomotor \cite{nanotubo}. In this article, we have shown the
conditions for which a nanowalker can work.

\begin{acknowledgments}
We thank L.~M. Flor\'{\i}a for helpful comments and discussion.  Work
is supported by the Spanish DGICYT Project FIS2005-00337.
\end{acknowledgments}


\begin{thebibliography}{99}
\bibitem{l0} Note that at the moment we are using the same constant
$l_0$ for the natural length of the neck, Eq.~(\ref{FENEeq}), and the
period of the microtubule lattice, Eq.~(\ref{period}). This constraint
will be relaxed later in section III-E.
\bibitem{com1} As the ATP gives to the head the energy needed to jump
over the potential barrier, the amplitude $V_0$ of the X-potential can
be taken to be the energy extracted from a single ATP molecule,
$E_{ATP}$ [Looking at \cite{biophysics}, at room temperature, $V_0
\equiv E_{ATP}\simeq 20k_B T=0.5\,eV$, therefore, $\widetilde{T}\simeq
0.05$]. As the external load $Q_0$ needed to make the system go
backward is known, it can be also used to fix $V_0$, as this force is
used in making the particle climb over the smoothest slope of the
potential considered. Therefore, using $x_M=0.5$, which makes the
slope considered be $1.5\cdot l_0$ meters width, $V_0=Q_0\cdot(1.5\,
l_0)=0.54\,eV$, and $\widetilde{T}=0.046$.  The external load needed to
stall the system and make it go backward has been experimentally
measured in \cite{Carter}, obtaining $Q_0=-7\,pN$.
\end{thebibliography}
\end{document}